# Clinical study:

# Two Optical Coherence Tomography Systems detect Topical Gold Nanoparticles in Hair follicles, Sweat Ducts and Measure Epidermis

*Mette Mogensen[1], Sophie Bojesen[1], Niels Israelsen[2], Michael Maria[2], Mikkel Jensen[2], Adrian Podoleanu[3], Ole Bang[2], Merete Hædersdal[1]*

1. Dept of Dermatology, Bispebjerg Hospital, University of Copenhagen, Bispebjerg Bakke 23, DK-2400 Copenhagen NV, Denmark
2. DTU Fotonik, Department of Photonics Engineering, Technical University of Denmark, Ørsteds Plads, building 343, DK- 2800 Kongens Lyngby, Denmark
3. School of Physical Sciences, Ingram Building, University of Kent, Canterbury, Kent CT2 7NH, UK

## Conflict of Interest:

None declared

## Corresponding author:

**Sophie Bojesen, MD**

Dept of Dermatology, Bispebjerg Hospital

University of Copenhagen

Bispebjerg Bakke 23

DK-2400 Copenhagen NV

Denmark

Email: sophiebojesen@gmail.com

Mobile: +45 27 583 556

Orchid ID: 0000-0002-7493-9665

# Introduction:

Optical coherence tomography (OCT) is an optical imaging technology that enables real time, high-resolution, cross-sectional and *en face* investigation of skin by detecting reflected broad-spectrum near-infrared light from tissue. OCT provides micron-scale spatial resolution and millimeter-scale depth of penetration [1]. Several commercial OCT systems with handheld probes targeted for Dermatology are now available [2].

The ability of OCT to achieve high diagnostic accuracy in skin diseases is hampered by the fact that not all diseases show sufficient contrast to be discriminated from normal skin. The challenge in realizing contrast enhancement in OCT imaging is to achieve signal from exogenous contrast agents that can overcome the intrinsic signal of the skin itself, allowing segregation of normal skin structures from skin pathology [3]. Our clinical experience, supported by animal studies [4;5], shows that application of a contrast agent, such as gold nanoshells (GNS), leads to a strong hyperreflective signal in OCT images deriving from the natural skin openings. This property of GNS can be used to generate enhanced contrast around hair follicles and sweat glands and potentially improve the diagnostic accuracy in some skin diseases.

The adnexal structures of skin consist of hairs, hair follicles, sebaceous glands, eccrine glands, and sweat ducts delivering the sweat to the skin surface. It is well-known that hair follicles are pathways from the upper skin layers to lower skin layers [6;7] and therefore contrast agents can be delivered through the pilous route.

The diagnostic accuracy of OCT in Dermatology has been demonstrated as moderate to high [8;9]. Several skin diseases have been studied clinically with OCT [10-15] including disorders involving adnexal structures: sweating [16], dyshidrotic skin conditions [17], follicular keratosis [18], abnormal hair growth [19], hair transplantation [20] and acne [21-23]. Identification and measurement of hair follicle calibers, size/condition of hair follicles and sweat gland structures can be useful in monitoring these diseases. The efficacy of OCT has already been established for investigation and measurement of hairs and nails [10;24].

Since OCT systems can be quite different in regards to light source, focus point, and data processing, the resolution and penetration depth varies among different systems. For example, data on thickness of keratinocyte carcinomas does not seem to correlate between OCT systems when compared clinically [25]. Several studies have compared different OCT systems used in dermatology. One study compared OCT imaging of actinic keratosis and basal cell carcinoma (BCC) in a clinical setting of 29 patients using three different OCT systems: VivoSight, by Michelson Diagnostics Ltd, UK; Callisto by Thorlabs Inc, USA; and Skintell by Agfa Healthcare NV, Belgium. A correlation between the first two devices regarding BCC tumor thickness was established, however, the BCC thickness did not correlate to histology in either. The latter OCT system Skintell showed no correlation with VivoSight and Callisto. Reassuringly, all OCT systems differentiated skin cancer from normal skin [25]. However, the discrepancy highlights the need to compare OCT systems head-to-head clinically.

The natural first steps of introducing topical GNS in the dermatological clinic is to describe their distribution in normal skin and in adnexal structures, imaged by more than one OCT system, and subsequently to apply GNS to diagnostic studies of the skin disease.

**Aim of study:** To describe and explore the utility of GNS as contrast agent for OCT imaging in normal healthy human skin with special regard to distribution in hair follicles and sweat ducts. Four anatomical areas of the skin were scanned: cheek, palm of the hand, armpit, and chest. To assess versatility of GNS and potential differences in OCT signal and OCT intensity the healthy participants were examined by two different OCT systems: Vivosight OCT (VS-OCT) that has a 5-7.5 µm resolution and a prototype ultra-high OCT (UHR-OCT) system with a 3-6 µm resolution. Furthermore, we compared the epidermal thickness measurements performed using the two OCT systems.

# Methods and Materials:

## Study design:

A clinical prospective study, consecutively including healthy volunteers from the Dept. of Dermatology at Bispebjerg Hospital, University of Copenhagen, Denmark was established in accordance with Helsinki II Declarations. The research protocol was approved by the Ethics Committee of The Capital Region of Denmark: no. H-16039077.

## Participants

Healthy participants were included from our department, from social media advertisement, approved by the Ethics Committee. Inclusion criteria were age above 18 and normal healthy skin. Exclusion criteria consisted of pregnancy or lactation, active skin disease or systemic disease requiring ongoing medical therapy. Participants were instructed to report any discomfort or skin irritation during and after the study.

### OCT systems:

The commercial OCT system is a VivoSight Rx OCT system (referred to here as VS-OCT, Michelson Diagnostics Ltd., Kent, UK) (. The VS-OCT is a multi-beam swept source frequency domain system with a tunable diode laser with a peak power of 15 mW at $\lambda$ = 1305 nm. Maximal field of image is 6 x 6 mm. It offers a lateral optical resolution of <7.5 µm and an axial resolution of 5 µm in skin. The penetration depth in skin varies around 1–2 mm and is limited by scattering effects. For each skin area of 4x4 mm, a cross-sectional multi-slice scan modality consisting of 250 B-scans was set up.

The ultra-high resolution OCT (UHR-OCT) is a prototype of a novel OCT system assembled at the Technical University of Denmark and NKT Photonics (Birkeroed, Denmark) with technical input from the Applied Optics Group, University of Kent, UK. The system uses a Supercontinuum light source combined with a broadband filter. The combination provides an output light beam with an average power of 5 mW on the skin and a wavelength range from 1000 nm to 1500 nm. The system delivers an axial resolution of 3 µm and a lateral resolution of 6 µm. A multi-slice scan modality of 1024 B-scans per 3x3 mm was set up for each skin area. Both OCT systems have the size of modern ultrasonography systems and handheld probes.

### Gold Nanoshell Contrast Agent:

GNS utilized have a 120 nm diameter silica core and 15 nm thick gold shell lining. These particular 150 nm GNS were selected because they have been demonstrated to penetrate into hair follicles easily and are approved for clinical use and for topical treatment of acne vulgaris combined with selective photothermolysis [26]. The microparticles are certified as medical equipment: CE 612960. The GNS used in this study are supplied in glass vials in quantities of 1 mL (Sebacia Microparticles, Sebacia Inc., Duluth, GA, USA). Microparticles are suspended in a combination of ethanol, polysorbate 80 and diisopropyl adipate. The gold particles come in a liquid dark blue solution that is to be applied to skin using a dispenser. Skin areas of 5 × 5 cm were cleansed and 0.25 mL of GNS suspension was massaged into each of four skin areas. Two stages of 30 seconds of massage with an oscillating handheld massage device (Sebacia Massager, Sebacia Inc., Duluth, GA, USA) was applied to each of four test areas.

### Clinical setting:

GNS was applied to cheek, palm of the hand, armpit and the front of the chest on each participant. Hairs were removed with a no-touch non-traumatic razor prior to OCT scanning. OCT images were acquired before and after GNS application. After application of GNS, skin was wiped with wet gauze. In one participant GNS was applied to lower arm and VS-OCT images were acquired after 20 min, 3 and 8 hours. Images of untreated skin and images of skin containing GNS contrast agent were recorded using both OCT systems. GNS penetration depth and epidermal thickness were assessed and compared in chest and armpit skin only. Morphological description of GNS in OCT images included all four skin areas. OCT scans were performed by authors SB, NI, and MeM.

### Comparison of pixel values:

In order to estimate intensity of gold particles in OCT images VS-OCT and UHR-OCT images from chest and armpit in all participants and the arm of subject 4 the 8-bit pixel values were counted before and after GNS application.

### Measurement of epidermal thickness (ET)

In obtaining the ET, a graphical approach was applied. For both OCT devices surface recognition was performed visually in OCT images using the integrated Vivosight software module and ImageJ (open source Java image processing program, NIH, 2017 online edition) 'Wand tracing tool' for VS-OCT and UHR-OCT images, respectively. By visual recognition of the skin surface in OCT images an identical line was used as an approximate delineation of the dermo-epidermal junction (DEJ) visually assessed and positioned by MeM. The axial distance between the surface line and the visually positioned ghost surface tracing DEJ constituted a final measure of the ET.

### Statistics:

Descriptive statistics were used and presented as means and medians with minimum and maximum ranges. Absolute and relative frequencies mean and standard deviation (SD) for continuous measurements were calculated for each parameter. Statistics are done using MATLAB 2017a (MathWorks® 2017 online version, Natick, MA, USA). Image data was analyzed with Wilcoxon paired test, as number of participants was deemed too small for assuming normal distribution. The Wilcoxon paired test was performed on the 8-bit pixel values and the pairs consist of one image from each system. The Wilcoxon test was performed with MATLAB *signrank*. A p-value < 0.05 was considered statistically significant. Bland-Altman plots were applied to detect a potential systematic difference between OCT measurements of epidermal thickness.

## Results:

A total of 11 participants were enrolled: 64% females, mean age 37.6 years with a median Fitzpatrick skin type of 2 (Table 1). GNS did not irritate the skin nor leave permanent discoloration. It was easily removed from skin with lukewarm tap water.

### Skin morphology in OCT images:

Before GNS application: as seen in Fig.1A and 1C, both OCT systems depict the distinct layering of epidermis, dermis and hairs. In OCT images epidermis is represented by an upper dark gray band, dermis below is lighter and less compact. Blood vessels are identified as black oval to round structures, in VS-OCT images blood vessels can be discriminated from lymphatic vessels by means of the integrated speckle variance OCT software (Dynamic OCT). Lymphatic vessels do not exhibit the high flow found in blood vessels and do not display a dynamic OCT signal. Hair follicles are identified as an oblique to vertical dark slender structure spanning through epidermis. Sometimes a hair is easily identified inside. In terminal hairs, such as beard hair in men and coarse hair in women, the bulge region is too deep to be imaged by OCT. The smaller vellus hairs present all over the body except from soles of hand and feet and mucosa are characterized by a dark bulge region in dermis next to sebaceous glands (Fig.1). The sweat ducts from the eccrine glands in the armpit, chest and face are not visible (Fig.2B, 2D) without GNS contrast agent, whereas sweat ducts are clearly visualized in the palm of the hand as distinct white hyperreflective spiral shaped structures inside the stratum corneum part of the epidermis (Fig. 3D). The eccrine glands themselves cannot be visualized, only the sweat ducts releasing the sweat produced onto the skin surface. Epidermal thickness in axillary skin tends to vary in thickness due to thickening around the terminal hair follicles (Fig. 2F). Epidermis is more homogenously dark without contrast agent in UHR-OCT images (Fig.2B, 2D, 2F).

After topical GNS application: Dilated serpiginous vessels characteristics for armpit skin were detected by both OCT systems and looked the same with and without GNS (Fig 2A and 2C). The GNS line the inside of hair follicles and occasionally cover the hair straw, generating an enhanced signal from the opening of the hair follicle (Fig.1B, 1D, 1E and 3A, 3C). Application of GNS improved contrast to normal skin in sweat glands of armpit skin and in hair follicles on the chest and armpits (Fig. 2E). Average penetration depths of 0.57 ± 0.17 mm in chest and 064 ± 0.14 mm in armpit were found. Maximum GNS penetration depth was 1.20 mm in hair follicles and only 15-40 µm in the sweat ducts of the axillary epidermis (Fig.1, table 1). In axillary skin, the sweat ducts was only visualized after application of gold as tiny white vertical strands in epidermis (Fig. 2E). The sweat ducts on the chest were barely visible (Fig. 1, asterix). GNS stick to skin surface increasing the entrance signal, which can impair OCT image quality (Fig. 3D), if not sufficiently wiped off. GNS penetrated deeper in armpit skin compared to the skin on the chest (Fig. 4A).

In Figure 3, VS-OCT images from all four anatomical regions demonstrate deposition of gold indicated by white arrows. The hair follicles that contain GNS exhibit various morphologies: bright cap on top or bright highly reflective dots from inside the hair follicle. On the palm, there is a thin bright upper band caused by GNS reflection. Some typical bright spiral shaped eccrine ducts are obscured by dark shadows after gold application.

All hairs are hyperreflective outside the hair follicle (Fig. 1), but black hairs are also hyperreflective inside the hair follicle (Appendix 1). There was no difference in intensity or overall skin morphology after topical application and massage of the GNS into lower arm of subject 4 after 3 and 8 hours. Gold was not detected in hair follicles at 8h, but at 20 min and 3h (Appendix 2). We did not find any contrast difference at the dermo-epidermal junction 3 or more hours after application of GNS (data not shown).

### Epidermal thickness (ET):
The ET appeared thinner in UHR-OCT images than in VS-OCT images from both armpit and chest skin (Fig. 4A). A statistically significant difference between ET measurements performed by the two OCT systems was demonstrated (chest $p< 0.039$, armpit $p< 0.027$). Bland-Altman plots were applied to detect a potential systematic difference between ET measurements using UHR-OCT and VS-OCT systems. The plots shown in Fig. 4B and 4C demonstrate that ET is slightly thinner in UHR-OCT, but OCT measurements of epidermis correlate and are not skewed in the respect that no OCT system differs more around certain values.

### Differences between UHR-OCT and VS-OCT:
The histograms in Fig.1 show OCT image pixel values of a hair follicle imaged by VS-OCT and UHR-OCT with and without gold. Fig. 1D shows the histograms for chest skin and Fig. 1E shows the histograms for axil skin. Without gold (dashed circles), the histograms look similar. After GNS application the pixel intensity was higher around hair follicles in UHR-OCT images compared to pixel values in VS-OCT images $p=0.002$ chest and $0.002$ armpit and it was further demonstrated that ET was statistically significantly thinner when measured with UHR-OCT (table 1 and Fig. 4A).

# Conclusion:

Application of 150 nm GNS improves contrast to normal skin in sweat glands in armpit skin and in hair follicles in general. GNS are highly scattering, independent of where they are located in skin. OCT detects signal from GNS presence in hair follicles, sweat ducts and demonstrate that GNS presents adhesion abilities on skin surface. Even though GNS was massaged into skin it was not detected by OCT beyond the depth of 1.2 mm into hair follicles and only 15-40 µm into sweat ducts of the axillary epidermis. The GNS penetrated deeper in armpit hair follicles than on chest. ET thickness was lower and pixel intensity higher in UHR-OCT system compared to the commercial system – however, both systems detected a significant enhancement of the OCT signal after application of GNS. Not only in hairs and sweat ducts, but contrast between epidermis and dermis was also slightly enhanced visually. GNS was not detected after 8 hours and did not irritate the skin nor leave discoloration.

# Discussion:

OCT can visualize skin layers, hairs and sweat ducts. Efficacy and limitations of OCT diagnosis of skin disease has previously been established in studies of skin appendages, such as hairs and nails [10;17;18;21;23;27-32]. However, various OCT systems with different axial and lateral resolution, different power on the skin, and different imaging post-processing software, have been applied in these studies. It is therefore imperative to compare commercially available and prototype OCT systems head-to-head, since the data retrieved from different OCT systems are not always comparable; i.e. in this study we find that epidermal thickness and pixel intensity generated from GNS differs between two OCT-systems. Differences are also demonstrated in another study were measurement of basal cell carcinoma thicknesses varies between three OCT-systems[27].

The statistically significant difference between ET measurements from two different OCT systems found here is clinically relevant. Measured ET values correlated within the two systems, but were significantly thicker in VS-OCT. We explain these differences as due to a higher resolution of the UHR-OCT system that enhances the features of the dermo-epidermal junction, thereby facilitating the visual segregating of epidermis from dermis. Future OCT studies must take into account that resolution does affect visual segregation of epidermis from dermis and thereby ET measurements.

Hair follicle and sweat gland duct contrast is apparently low in some skin regions such as cheek, chest and armpit. Utilization of topically applied contrast-enhancing agents such as GNS may improve sensitivity of OCT skin investigation. Application of GNS as contrast agent increased pixel intensity significantly, especially around hair follicles and with a higher intensity in the UHR-OCT system compared to VS-OCT.

This is the first time that topically applied GNS has been investigated with OCT *in vivo* on human skin, but previous *in vivo* animal studies have been performed: a rabbit skin study [5] applied silica–GNS with 150 nm silica core size and 25 nm lining. An increased intensity of the OCT signal in areas containing GNS was detected. GNS caused increased brightness in the superficial part of the dermis and higher contrast between the superficial and deep parts of the rabbit dermis already after 30 min. After 3 h, the changes in OCT images became more pronounced. Contrast was enhanced in hair follicles and sweat glands in OCT images, the effects lasted up to 24 h. The penetration of the 200 nm diameter GNS into the upper part of the rabbit skin (epidermis, upper part of dermis) was confirmed by electron microscopy. Three hours after application, GNS were rarely seen in epidermis, but detected in dermis: in fibroblasts and among collagen fibers, inside cells, and in the intercellular substance. Another OCT study performed on on pig skin [4] also demonstrated an increased contrast in OCT images after a single application of GNS with 75 nm silica core and 25 nm gold lining. Multiple applications of gold did not increase the contrast. Monte Carlo calculations were performed and simulations exhibited good qualitative agreement with the experimental images, and proved that GNS were responsible for the enhanced contrast. The OCT images collected 150 min after applying GNS demonstrated an OCT signal, which considerably exceeded the reference image. This suggests that the 125 nm diameter GNS penetrated into the pigskin and provided increased backscattering at all depths.

In the referenced animal studies, penetration of GNS into the dermis was observed. We did not detect GNSs in dermis and we suggest that the epilation process must have caused epidermal damage in the animal studies, thus allowing GNS to enter dermis. None of the referenced studies reported their hair removal methods. We only detected gold accumulation through natural skin openings, with an affinity to hair follicles. Furthermore, we used a massage device on intact human skin. We did not find any difference in contrast intensity at the dermo-epidermal junction 3 or more hours after application of GNS. Our findings are supported by a review of literature reporting that GNS larger than 100 nm do not penetrate the intact skin barrier [33]. The handheld massage device applied to intact skin in this study thus gently forced GNS into hair follicles and sweat ducts. We could have chosen GNS with even smaller diameter than 150 nm, since the pig skin study showed good penetration of 125 nm GNS into deeper layers, but so did the rabbit skin study that applied 200 nm diameter GNS. Therefore in this study we chose a GNS diameter in between.

Since GNS are potential nanotechnology-based drug delivery systems that can accommodate one or more active drugs of nanosize to be dispersed, absorbed, conjugated or encapsulated - GNSs can be targeted to be released in cancer or specific organs. Several experimental studies have implemented GNS in drug delivery systems e.g. by attaching GNS to anti-cancer antibodies [34-36].

Conclusively, this clinical study provides novel data on GNS used as a physical contrast agent. GNS is easy to use in a clinical setting and can be topically applied to skin generating increased OCT signals from natural skin openings. This study suggests that GNSs are interesting candidates for increasing sensitivity in OCT diagnosis of hair and sweat gland disease because GNS primarily enhances OCT contrast in hair follicles and sweat ducts. This should encourage future trials exploring not only diagnostic accuracy of GNS in OCT imaging but also look into GNS targeted drug delivery monitored by OCT.

# Supporting information

**Acknowledgements:**

We would like to acknowledge support from Innovation Fund Denmark through the ShapeOCT grant No. 4107-00011A. O. Bang and M. Maria acknowledge support from European Union's Horizon 2020 grant GALAHAD (732613). Additionally, M. Maria, A. Podoleanu, and O. Bang acknowledge the UBAPHODESA Marie Curie European Industrial Doctorate. A. Podoleanu is also supported by the Royal Society Wolfson Research Merit Award and the NIHR Biomedical Research Centre (BRC) at Moorfields Eye Hospital NHS Foundation Trust, UCL Institute of Ophthalmology.

# Legends:

**Figures:**

Figure 1: **Two OCT systems: skin, hairs and pixel values in a hair follicle with and without 150 nm gold nanoshells (GNS)**

OCT images from VS-OCT and UHR-OCT. Images captured from chest skin of subject 6. The hair follicle without contrast agent is an oblique dark-grey elongated structure penetrating epidermis from dermis. The hair straw is not visible without contrast agent. After application of GNS the hair straw is enhanced and appears as a bright slender structure inside the hair follicle. In UHR-OCT the sweat ducts are also vaguely visualized as a grainy white speckle inside epidermis (white asterix), epidermis is more homogenously dark without contrast agent. A 3D image of a hair on the cheek from the same participant show affinity of the gold to the hair straw outside the skin and the gold lining the hair follicle inside the skin.

Histograms of pixel values in chest and armpit skin. Without GNS (dashed circles), the histograms for the two systems are similar. Application of GNS increased pixel intensity significantly, especially around hair follicles and with a higher intensity in the UHR-OCT system compared to VS-OCT. No increase in the highest pixel values was detected in VS-OCT after application of GNS, but rather a shift of the statistics towards lower pixel values (Fig. 1E). This inclination indicates that a normalization of the pixel value distribution is performed according to the absolute signal values detected.

Figure 2. **OCT-images from the armpit skin demonstrating epidermal thickness (ET) and sweat ducts with and without 150 nm gold nanoshells (GNS)** All ETs are marked by blue bar. 2A: VS-OCT system demonstrates a dark-grey epidermal band with fluctuating thickness. Dilated serpiginous vessels characteristics for armpit skin marked by rings. No GNS is identified in sweat ducts nor in hair follicles. Contrast between epidermis and dermis is slightly enhanced. 2B: VS-OCT system. No GNS were applied. GNS not identified neither in sweat ducts nor in hair follicles. 2C: UHR-OCT, similar to 2A serpiginous dilated vessels is marked by a circle. 2D: UHR- OCT, no GNS applied. Contrast between epidermis and dermis is less visible than in 2C. 2E: UHR-OCT, sweat ducts stand out as thin white hyperreflective vertical strings spanning from top of the skin into epidermis. 2F: no increased signal identified neither in sweat ducts nor in hair follicles. 2G: VS- OCT system. GNS identified in 3 hair follicles but not in sweat ducts. The hair follicle far right

contains a bright grain of GNS identified by arrow. 2H: VS-OCT system, the black round area is a very superficial blood vessel.

Figure 3: **OCT images of healthy volunteers after topical delivery of gold nanoshell (GNS) as contrast agent.** OCT images of anatomical skin regions included in the study after topical delivery of GNS. Deposition of GNS is marked by white arrows: In 3A and 3C, the GNS are deposited inside the hair follicle, in 3B the hair follicles containing GNS have various morphologies: bright white cap on top of the skin and bright hazy highly refractive signals from inside hair follicle. On the palm, 3D, a bright band on top of the skin is caused by surface accumulation of GNS. The typical bright spiral shaped eccrine ducts (white asterix) are obscured by dark vertical shadows caused by gold deposition in the opening of the sweat duct.

Figure 4: **Epidermal thickness (ET) and penetration depth of gold nanoshells (GNs) in OCT images from armpit and the chest.** Top: Penetration of GNS into skin measured in OCT images from the armpit and chest of healthy participant 2-11 (black and grey color). GNS penetrate deeper in armpit, also termed axilla. Bottom: Means of ET (yellow and blue) measured in OCT images from Participant 1-11 in both skin areas. 4B and 4C are Bland-Altman plots demonstrating no systematic skewedness of ET measurements when UHR-OCT is compared to VS-OCT in chest and armpit, respectively.

## Tables:

Table 1: **Data from Healthy Participants.** Data from all eleven subjects included. Penetration depth of GNS into skin is in bold. Descriptive statistics of epidermal thickness included all 11 volunteers, but subject 1 was excluded from the measurement of GNS penetration depth as she had 2 ml of gold contrast agent applied to skin instead of 1 ml.

**Appendix Legends**

Figure Appendix 1: **Dark hairs versus gold nanoshells (GNS) in hair follicles.** OCT images of a dark hair in the beard region shows that black hairs can be traced into the dermis and appear as thin hyperreflective rods inside dermis. After topical GNS application the gold particles tend to form dots and caps inside of dermis. Hence the morphology of black hairs can be differentiated from GNS lining the hair follicle and hair itself.

Figure Appendix 2: **Consecutive OCT images from the VS-OCT system over time after application of 150 nm gold nanoshells (GNS) at timepoint: 20 min, 4h, 8 h.** Three VS-OCT images of a small light brown dermal nevus on the lower left arm of Participant 4. In A: 20 min after application of GNS, a clinical photo of the nevus is inserted. The nevus is well-circumscribed oval dark grey, marked by "N". In B, at 4h the outline of the nevus and the penetration depth is unchanged. GNS are detected in an adjacent hair follicle (arrow). In C, GNS are located in the middle of the nevus, very superficially after 8h (arrow). Due to strong reflectance from GNS at the nevus surface, a shadow is created below the GNS contrast agent. In images A-C, a strong entrance signal is detected on top of the skin, even though GNS has been wiped off.

| Healthy Volunteer no. | Age | Fitz-patrick skin type | Sex (female/male) | Penetration depth gold nanoshells, chest (mm) | Penetration depth gold nanoshells, armpit (mm) | Epidermal thickness, chest. UHR-OCT (mm) | Epidermal thickness, chest. Vivosight (mm) | Epidermal thickness, armpit UHR-OCT (mm) | Epidermal thickness, armpit Vivosight (mm) |
|---|---|---|---|---|---|---|---|---|---|
| 1 | 33 | 3 | f | **0.66** | **0.60** | 0.09 | 0.10 | 0.12 | 0.17 |
| 2 | 30 | 2 | m | **0.60** | **0.51** | 0.08 | 0.10 | 0.13 | 0.16 |
| 3 | 45 | 3 | f | **0.38** | **0.76** | 0.08 | 0.10 | 0.12 | 0.17 |
| 4 | 27 | 2 | f | **0.45** | **0.52** | 0.07 | 0.11 | 0.14 | 0.12 |
| 5 | 43 | 2 | f | **0.55** | **0.71** | 0.08 | 0.10 | 0.13 | 0.12 |
| 6 | 30 | 2 | m | **0.68** | **0.64** | 0.10 | 0.13 | 0.13 | 0.19 |
| 7 | 72 | 2 | f | **0.79** | **0.94** | 0.09 | 0.11 | 0.13 | 0.16 |
| 8 | 51 | 2 | f | **0.30** | **0.53** | 0.11 | 0.15 | 0.19 | 0.17 |
| 9 | 28 | 2 | m | **0.41** | **0.60** | 0.08 | 0.11 | 0.15 | 0.16 |
| 10 | 28 | 3 | m | **0.67** | **0.70** | 0.09 | 0.11 | 0.17 | 0.17 |
| 11 | 27 | 2 | f | **0.82** | **0.49** | 0.09 | 0.08 | 0.16 | 0.22 |
| Results: Mean and (95% CI) | mean = 37.6 | median = 2 | Female = 64% | **0.57** SD 0.17 | **0,64** SD 0.14 | 0.09 mm (0.07, 0.11) | 0.11 mm (0.07, 0.15) | 0.14 mm (0.10, 0.18) | 0.16 mm (0.11, 0.22) |
| **p-values:** | | | | | | **Chest:** | **0.039** | **Armpit:** | **0.027** |

Table 1: **Data from Healthy Volunteers**

# Figures

Figure 1:

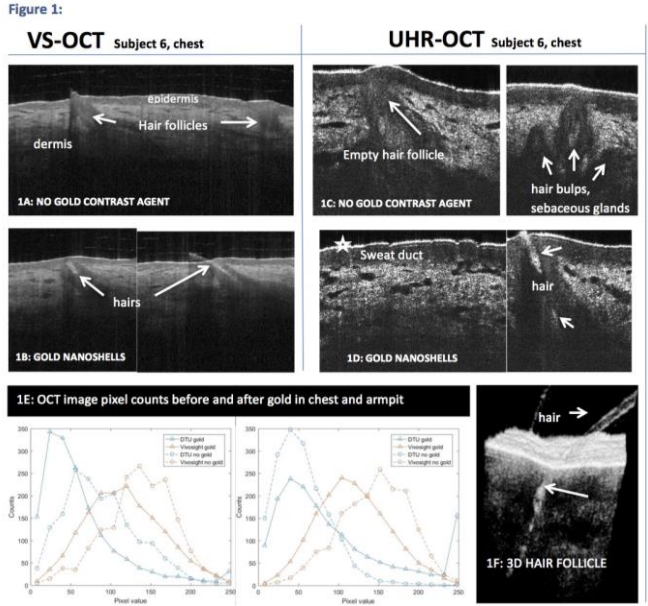

Figure 2:

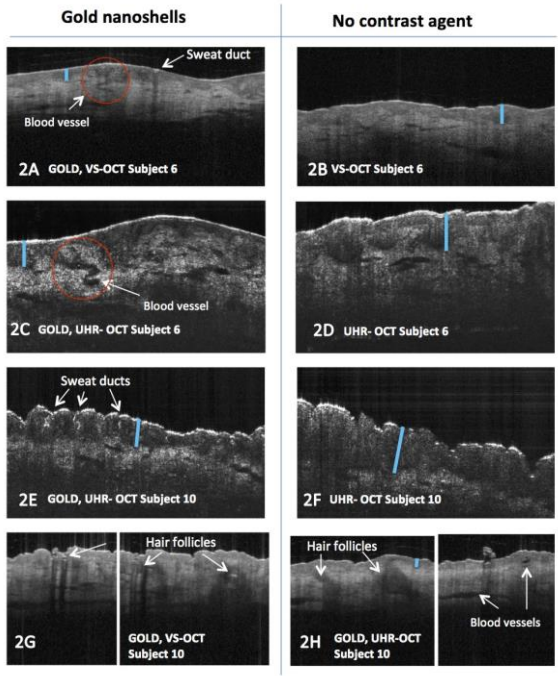

Figure 3:

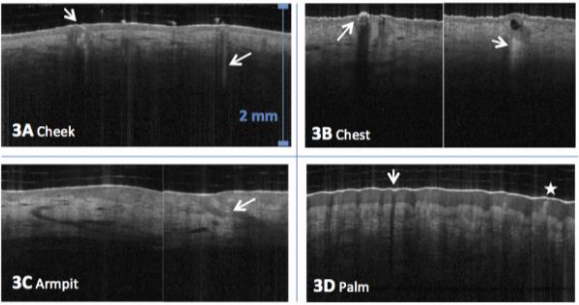

Figure 4: **Epidermal thickness and penetration depth of gold nanoshells in OCT images from armpit and chest**

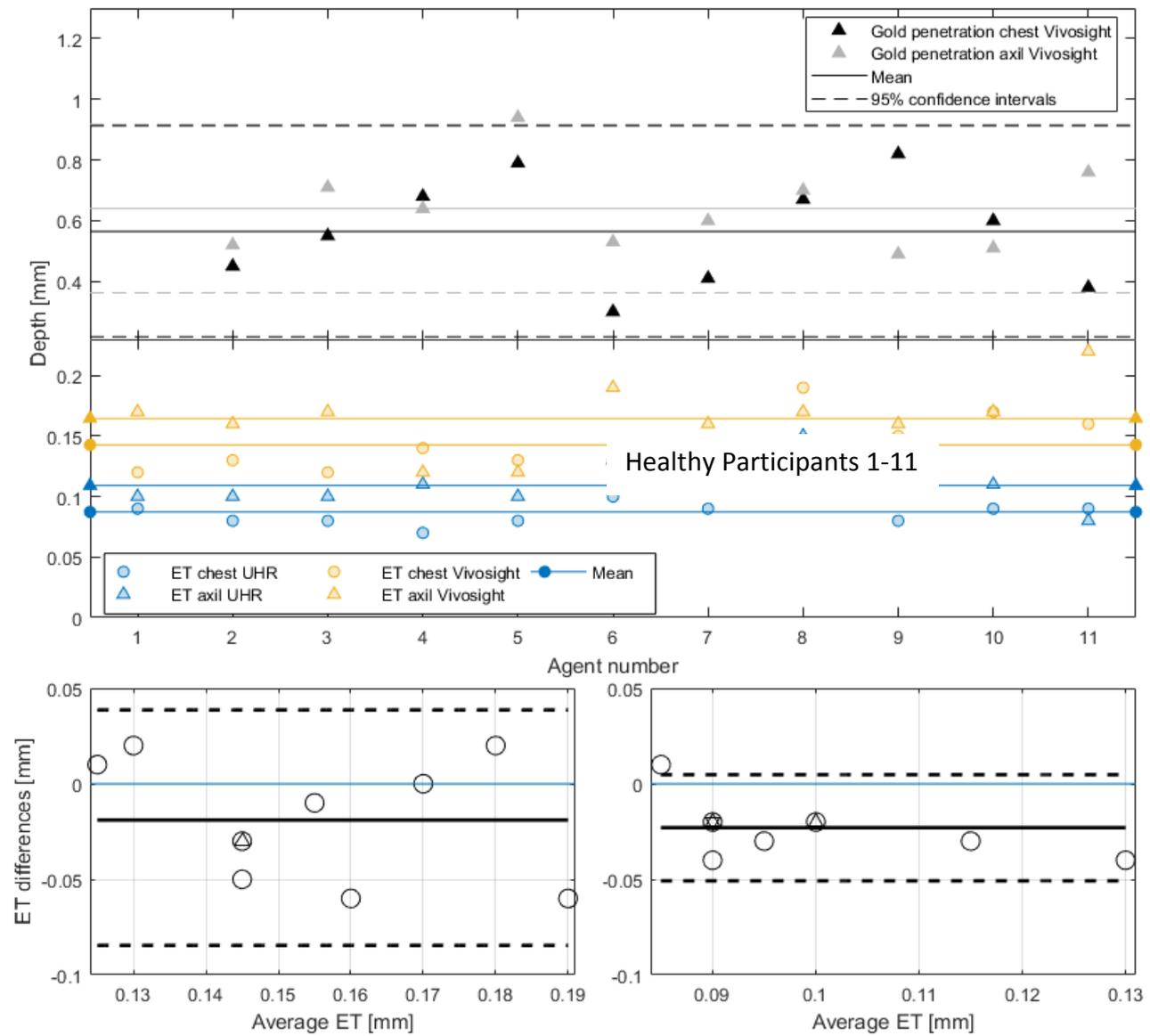

4A

4B   4C

# Figure Appendix 1

**A1: VS-OCT demonstrated dark hairs versus gold in hair follicles.**
Subject 8, cheek

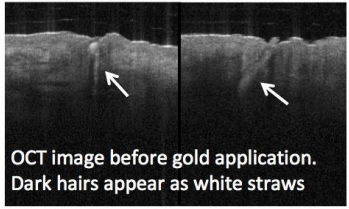
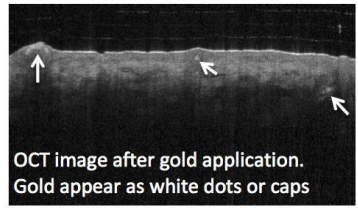

OCT image before gold application. Dark hairs appear as white straws

OCT image after gold application. Gold appear as white dots or caps

# Figure Appendix 2

Appendix 2

Consecutive OCT images from the Vivosight system over time after application of 150 nm gold nanoshells at timepoint: 20 min, 4h, 8 h.

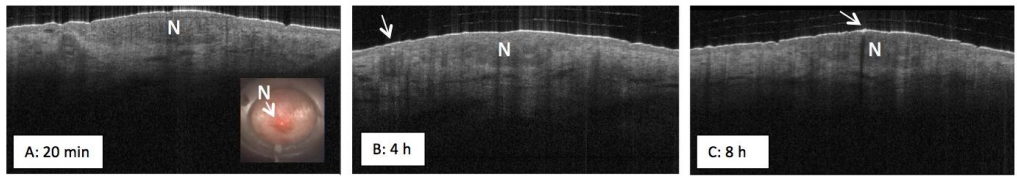

A: 20 min  B: 4 h  C: 8 h